\documentclass{iopart}
\usepackage{epsfig}

\begin{document}

\jl{4}

\title{Centrality dependence of $K^+$ produced in Pb+Pb collisions
 at 158 GeV per nucleon}\footnote{Work supported in part by the Schweizerischer
Nationalfonds.}

\author{
Sonja Kabana for the NA52 collaboration:\\
\vspace*{0.2cm}
\normalfont 
\small
G.~Ambrosini$^1$, R.~Arsenescu$^1$, C.~Baglin$^3$, H~P~Beck$^1$, 
K.~Borer$^1$, A.~Bussi\`ere$^3$,
K.~Elsener$^2$,
Ph.~Gorodetzky$^5$, J.P.~Guillaud$^3$, P.~Hess$^1$,
 S.~Kabana$^1$, R.~Klingenberg$^1$, G~Lehmann$^1$,
 T.~Lind\'en$^4$, K.D.~Lohmann$^2$,
R.~Mommsen$^1$, U.~Moser$^1$,  K.~Pretzl$^1$, J.~Schacher$^1$,
  F.~Stoffel$^1$, R~Spiwoks$^1$, J.~Tuominiemi$^4$, M.~Weber$^1$
}

\address{$^1$\ Laboratory for High Energy Physics, University of Bern,
    Sidlerstrasse 5, CH-3012 Bern, Switzerland, }
\address{$^2$\ CERN, CH-1211 Geneva 23, Switzerland, }
\address{$^3$\ CNRS-IN2P3, LAPP Annecy, F-74941 Annecy-le-Vieux, France, }
\address{$^4$\ Dept. of Physics and Helsinki Institute of Physics, 
 University of Helsinki,
    PO Box 9, FIN-00014 Helsinki, Finland, }
\address{$^5$\ PCC-College de France, 11 place Marcelin Berthelot, 75005 Paris, France}

\begin{center}

\begin{abstract}

\noindent
The NA52 collaboration searches for a
 discontinuous behaviour of charged kaons produced
in Pb+Pb collisions at 158 A GeV as a function of the impact parameter,
which could reveal a QGP phase transition.
The $K+$ yield is found to grow
 proportional to the number of participating ('wounded') nucleons
N, above N=100.
Previous NA52 data agree with the above finding
and show a discontinuous behaviour in the kaon
centrality dependence near N=100, 
marking the onset of strangeness enhancement -over e.g. p+A data
at the same $\sqrt{s}$-
 in a chemically equilibrated phase.

\end{abstract}

\end{center}

\begin{figure}[t]
\hspace*{-0.2cm}
\hspace*{-0.2cm}
\begin{center}
 \epsfig{figure=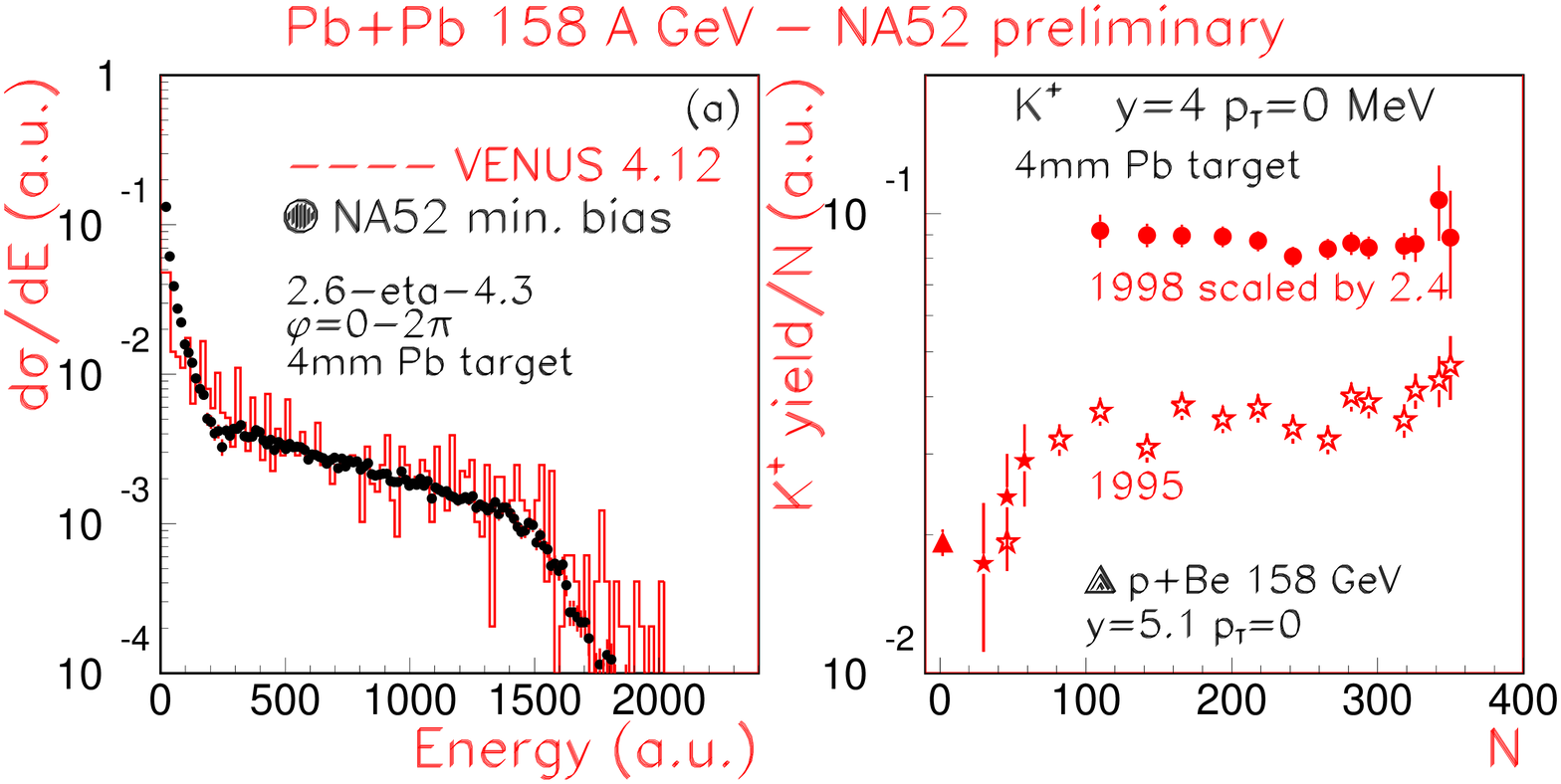, width=135mm, height=55mm}
\hspace*{-0.2cm}
\caption{
 Preliminary NA52 data from Pb+Pb collisions at 158 A GeV.
(a) Calorimetric energy distribution in arbitrary units.
(b):  $K^+$ yield in arb. units per N, as a function  of
N.
For comparison the 1995 NA52 data \protect\cite{na52_centr} are also
shown. 
The 1998 data are scaled by 2.4 with respect to the 1995 data.
}
\end{center}
\hspace*{-0.2cm}
\label{kaons}
\hspace*{-0.2cm}
\hspace*{-0.2cm}
\end{figure}

\section{Introduction}

\noindent
The quark-gluon plasma phase transition predicted by QCD \cite{qcd}
 may occur and manifest itself in ultrarelativistic heavy ion collisions
 through discontinuities in the temperature and
energy density dependence of relevant observables.
A major 
example of such a discontinuity is seen in the $J/\Psi$ to Drell Yan ratio 
\cite{na50_jpsi}.
The NA52 collaboration searches for discontinuities in strangeness
production measuring charged kaons 
as a function of the impact parameter.
 Results from the 1995 NA52 run are published in \cite{na52_centr}.
We report here on new preliminary results
from the 1998 run of the NA52 experiment, 
on $K^+$ at rapidity 4.1 and transverse momentum near 0
produced in Pb+Pb collisions at 158 A GeV.
In this run
a new electromagnetic lead/quartz fiber  calorimeter (QFC) 
with improved acceptance and resolution \cite{qfc}
was used.

\section{Results and discussion}

\noindent
For the kaon measurement 
we modified the 1998 set up of NA52 \cite{michele}
 by placing the target 0.6 m upstream of the calorimeter.
The results 
have been corrected for empty target contributions.
The number of participant nucleons N has been estimated from
the energy measured with the calorimeter (figure 1, (a)) in the way
described in \cite{na52_centr}. 
Particle identification is described in \cite{na52_centr}
and references there.
\noindent 
The positive kaon yield divided by N
is independent of N, for N $>$ 100 (figure 1, (b))
in agreement with previous NA52 results \cite{na52_centr}.
Assuming that N is proportional to the volume of the particle source,
figure 1, (b)
 shows that the kaon number density 
exhibit a discontinuity, saturating  above N=100.
This  indicates a transition to a phase characterized by a high degree of 
chemical equilibrium and enhancement \cite{na52_centr}
of kaons from the point N=100 on, corresponding to energy density
 $\epsilon$ $\sim$ 1.3 GeV/fm$^3$ \cite{hepph_0004138}, near
$\epsilon_c$ \cite{qcd,satz_review}.
This change marks the onset of strangeness enhancement seen in kaons
 in an equilibrated phase,
which may be suggestive for a QCD phase transition,
depending on the simultaneous appearance of thresholds in other signatures 
 at the relevant $\epsilon$ values and their theoretical understanding
  \cite{satz_review,na50_jpsi,hepph_0004138}.


\section*{References}
\begin{thebibliography}{9}
\bibitem{qcd} E. Laermann, Nucl. Phys. A610 (1996) 1c.
\bibitem{na50_jpsi} M.C. Abreu et al. (NA50 coll.), CERN-EP-2000-013.
\bibitem{na52_centr}
G. Ambrosini et al. (NA52 Coll.), N. J. of Phys. (1999) 1, 22,
 G. Ambrosini et al. (NA52 Coll.), N. J. of Phys. (1999) 1, 23,
S. Kabana et al. (NA52 Coll.), Nucl. Phys. A 661 (99) 370c.
\bibitem{qfc}
M. Weber et al (NA52 Coll.), CALOR97, World Scientific (1998) 151.
\bibitem{michele}
M. Weber et al (NA52 Coll.), these proceedings and Un. of Bern preprint
BUHE-00-04.
\bibitem{hepph_0004138} S. Kabana, hep-ph-0004138 and hep-ph/0010228. 
\bibitem{satz_review} H. Satz, hep-ph/0007069, P.B. Munzinger, 
J. Stachel, nucl-th/0007059.
\end{thebibliography}
\end{document}